\documentclass[12pt]{article}
\usepackage[utf8]{inputenc}
\usepackage{multirow}
\author{Author Name}
\DeclareUnicodeCharacter{03A6}
{\ensuremath{\Phi}}

\usepackage{amsmath}

\usepackage{amssymb}
\usepackage{tabularx}
\usepackage{newunicodechar}
\usepackage{siunitx}
\newunicodechar{ }{~}

\usepackage{doi}

\usepackage{longtable}

\pdfoutput=1
\usepackage{graphicx}

\usepackage{float} 

\usepackage{cite}
\usepackage{tabularx}

\usepackage{microtype} 

\usepackage[utf8]{inputenc}
\usepackage{longtable}
\usepackage{array}
\usepackage{ragged2e}
\usepackage[utf8]{inputenc}
\usepackage{enumitem}
\usepackage{xurl}

\usepackage{float}

\title{From Cell Towers to Satellites: A 2040 Blueprint for Urban-Grade Direct-to-Device Mobile Networks}
\author{Sebastian Barros}
\date{July 12th , 2025}  

\begin{document}
\renewcommand{\arraystretch}{1.2}

\maketitle

\begin{abstract}
In 2023, satellite and mobile networks crossed a historic threshold: standard smartphones, using unmodified 3GPP protocols, connected directly to low Earth orbit (LEO) satellites. This first wave of direct-to-device (D2D) demonstrations validated the physical feasibility of satellite-based mobile access. However, these systems remain fallback-grade: rural-only, bandwidth-limited, and fully dependent on Earth-based mobile cores for identity, session, and policy control.

This paper asks a more ambitious question: Can a complete mobile network, including radio access, core functions, traffic routing, and content delivery, operate entirely from orbit? And can it deliver sustained, urban-grade service in the world’s densest cities?

We present the first end-to-end system architecture for a fully orbital telco, integrating electronically steered phased arrays with 1000+ beam capacity, space-based deployment of 5G core functions (UPF, AMF), and inter-satellite laser mesh backhaul. We analyze spectral efficiency, beam capacity, and link budgets under dense urban conditions, accounting for path loss, Doppler, and multipath. Simulations show that rooftop and line-of-sight users can sustain 64-QAM throughput, while street-level access is feasible with relay or assisted beam modes.

The paper outlines the remaining constraints, power, thermal dissipation, compute radiation hardening, and regulatory models, and demonstrates that these are engineering bottlenecks, not physical limits. Finally, we propose a staged 15-year roadmap from today’s fallback D2D systems to autonomous orbital overlays delivering 50–100 Mbps to handhelds in megacities, with zero reliance on terrestrial infrastructure.
\end{abstract}

\section{Introduction}

The architecture of mobile networks has remained fundamentally Earth-bound for over half a century. While radio access has evolved from second to fifth generation, and core networks have moved toward virtualization and edge computing, one assumption has held constant. All critical infrastructure, including antennas, base stations, and core functions, remains on the ground.

Satellite communications have followed a parallel path. From early passive reflectors in the 1960s to current low Earth orbit broadband constellations, satellites have supported broadcasting, remote access, and specialized communication services. These systems have traditionally operated outside terrestrial mobile frameworks, requiring dedicated terminals, nonstandard protocols, and isolated spectrum.

In 2023, a new threshold was crossed. AST SpaceMobile and Lynk Global successfully connected standard smartphones to satellites in low Earth orbit using existing mobile frequency bands. These first demonstrations of direct-to-device connectivity from space used off-the-shelf devices and three GPP waveforms, with no additional hardware. However, these implementations remain constrained. They offer low data rates, limited concurrency, rural-only coverage, and depend entirely on terrestrial core infrastructure for user identity, session control, and policy enforcement.

This paper asks a fundamental question. Is it possible to operate a complete mobile network entirely from space, including the radio access layer, core network functions, authentication, mobility management, and application delivery, while providing reliable broadband service to users in dense urban environments?

This is not a trivial extension of rural connectivity models. Urban deployment presents the hardest case. Non-line-of-sight conditions, heavy multipath, high Doppler shift, building penetration loss, and dense spectral reuse requirements make space-based access extremely challenging. At the same time, orbital systems face strict constraints in power availability, thermal dissipation, and real-time computing.

We present a full system architecture for a mobile network deployed entirely in low Earth orbit. This includes electronically steered multi-beam antenna arrays, satellite-based radio access functions, partial core functions such as user plane and access management, inter-satellite optical routing, and onboard application edge nodes. The system supports both direct-to-device links and relay-assisted coverage in city environments with minimal reliance on ground infrastructure.

This paper makes four key contributions

 \begin{enumerate}
     \item We model the urban link budget for direct satellite to device connectivity using realistic path loss profiles, beamforming gains, and Doppler conditions across multiple bands.
     \item We simulate the spectral efficiency, signal quality, and user capacity per satellite beam under practical array size and power constraints.
     \item We define an architecture for virtualized core network functions operating in orbit, including access control, session management, and traffic breakout through optical mesh backhaul.
     \item We identify critical technical bottlenecks such as session continuity, thermal limits, orbital compute, and radio interference, and outline a development roadmap to achieve orbital mobile service in dense areas by 2040.

 \end{enumerate}
Our findings indicate that an orbital mobile network is feasible in dense urban settings. No physical law prevents this. The limiting factors are engineering complexity, system integration, and spectrum governance. The next generation of mobile infrastructure may not be terrestrial. It may reside in orbit, be stateless by design, and operate with global coverage by default.

\section{Background}

\subsection{Historical Evolution of Satellite and Mobile Networks}

Project Echo, launched in 1960, was the first passive communications satellite and used large balloon reflectors to bounce microwave signals between ground stations \cite{echo,echo1}. 
Its successor Echo 2, launched in 1964, followed the same concept and remained operational until its deorbit in 1969 \cite{echo}. 
These early experiments proved the concept of satellite as a long-range relay link.

The first geostationary communications satellite, Syncom 3, was launched in August 1964. 
It relayed live television from Tokyo to the United States during the 1964 Olympics \cite{syncom,syncom2025}. 
This milestone established continuous broadcast capabilities using a fixed orbital platform.

Terrestrial mobile networks evolved separately. 
First-generation analog systems began in the 1970s, followed by second-generation digital standards in the 1990s, ushering in efficient cellular reuse, mobility, and low-latency voice and data \cite{satcom}.

For decades, satellite and cellular networks operated in separate verticals: satellites for broadcast and remote connectivity, and cellular systems for dense urban access and mobility. 
No convergence occurred at the access or core layers until the recent emergence of low Earth orbit constellations.

This historical divide explains why, until now, satellites have not been integrated directly into standard mobile infrastructure.

\subsection{Milestone in Direct-to-Device Satellite Connectivity}

In 2023, AST SpaceMobile achieved the first successful space‑based voice call and data session using an unmodified smartphone and LTE or 5G waveforms. Their BlueWalker‑3 test satellite connected a standard Galaxy S22 to AT\&T spectrum, achieving download speeds over ten megabits per second during a demo in Hawaii, and later demonstrating the first 5G call capability using AT\&T spectrum in September \cite{ast4g,ast5g}.

Lynk Global followed with its demonstration of satellite-to-cell-phone connections. In July, two‑way voice calls were completed using standard handsets and Lynk satellites, marking the first public demonstration of direct-to-device voice from space \cite{lynk}. Licenses filed with the FCC and agreements with mobile operators like Turkcell and MTN reaffirm commercial sanction for this approach \cite{turkcell,mtn}.

While groundbreaking, these systems remain limited to low-throughput fallback services in rural areas. Typical data rates remain in the single-digit megabits per second range, beam count is low, coverage depends on a clear line of sight, and core network routing remains anchored on Earth.

\subsection{Three GPP Non-Terrestrial Network Architecture}

In Release 17, Third Generation Partnership Project added support for Non-Terrestrial Networks to its 5G New Radio standard, enabling satellites and aerial platforms to act as mobile access points \cite{3gpp_ntn}.

The architecture specifies multiple integration modes: transparent relay, regenerative payloads performing physical layer functions, and integrated nodes capable of executing full radio access logic onboard \cite{3gpp_ntn, ntn_simplified}.

Despite these enhancements, all terrestrial implementations continue to host core functions like access management, session control, and user identity on ground infrastructure. Satellites currently act primarily as radio relays with limited signal processing \cite{ntn_isolated}.

The specifications also include mechanisms to address challenges unique to satellite channels, such as Doppler shift, timing uncertainties, random access tuning, and extended latency. However, they do not define scenarios where all core network functions reside in space or control is fully decentralized \cite{3gpp_ntn}.

\subsection{How Satellite to Device Works Today}

Today, the most advanced implementations of satellite-to-device connectivity are built around large phased array antennas deployed on low Earth orbit satellites operating in licensed mobile bands. These systems aim to allow everyday smartphones to connect directly to space-based towers using standard 3GPP protocols, with no hardware modification. Unlike traditional satellite phones, which rely on proprietary handsets and protocols, direct-to-device systems emulate terrestrial base stations in orbit. This requires high-gain antennas, precise beam control, and full compliance with mobile signaling stacks.

AST SpaceMobile’s BlueWalker 3, launched in late 2022, provides a reference architecture. The satellite carries a planar phased array antenna that unfolds to over 64 square meters in orbit, making it the largest commercial phased array ever deployed in space. This array consists of thousands of radiating elements arranged in a rectangular lattice. Each element can adjust its transmit and receive phase, enabling the formation of narrow, high-gain radio beams that can be steered electronically toward specific ground locations. By concentrating power in a tight spatial footprint, the array compensates for the extreme free-space path loss encountered over LEO distances. From an altitude of roughly 500 kilometers, the path loss exceeds 160 dB at 2 GHz, demanding highly directional beams to maintain a viable link budget.

On the downlink, the satellite transmits LTE or 5G New Radio signals using baseband waveforms that are compliant with 3GPP standards. The phone receives these signals as if they were coming from a conventional terrestrial base station. Synchronization signals, cell IDs, and timing advances are processed by the phone’s chipset without any special firmware. The phone initiates a connection using standard random access procedures. Uplink transmission from the phone, however, presents a greater technical challenge. Smartphones transmit at relatively low power,  typically between 0 and 23 dBm, and this signal is severely attenuated by the time it reaches the satellite. To overcome this, the phased array applies receive beamforming, combining the weak signal coherently across many elements. This technique provides array gain typically in the range of 20 to 30 dB, sufficient to lift the uplink above the receiver noise floor.

Doppler shift is another critical factor. A LEO satellite travels at roughly 7.5 kilometers per second, resulting in Doppler shifts of up to 100 kHz in the S-band. To maintain synchronization, the satellite performs real-time frequency offset correction using software-defined radio and FPGA-based processing. It continuously tracks the relative velocity vector and adjusts both the frequency and timing alignment of the transmitted signal. This is essential for maintaining a stable connection during the short window, typically 3 to 5 minutes, when the satellite is assigned to a given ground user.

Spectrum access is handled through coordination with terrestrial mobile operators. For instance, AST SpaceMobile has partnered with AT\&T to use licensed S-band and PCS spectrum under roaming agreements. This ensures regulatory compliance and allows user devices to operate under their existing SIM and subscription. Lynk Global, another major actor in this domain, operates in lower frequency bands such as 617 to 960 MHz. These lower bands are advantageous for building penetration and link robustness, though they offer less available bandwidth than midband allocations. Lynk satellites are smaller and rely on denser constellations to provide overlapping coverage with shorter revisit times.

While the physical layer and radio interface emulate terrestrial networks, the logical architecture remains grounded. All authentication, session control, mobility management, and billing operations are handled by the terrestrial core network. The satellite functions effectively as a remote radio head, forwarding bearer and control plane traffic to Earth via high-capacity RF or optical links. Some systems use optical inter-satellite links (OISLs) to relay data across the constellation before routing it down to a terrestrial gateway near the user’s home network.

Data rates achieved in public tests have reached 10 Mbps under optimal rural conditions with low interference. These figures are orders of magnitude lower than terrestrial base stations but demonstrate that direct-to-device is technically viable for basic connectivity, messaging, and even voice calls. Both AST and Lynk have successfully demonstrated two-way voice service to standard smartphones, with AST claiming the first satellite-based 5G call using a commercial Samsung Galaxy S22.

Despite these breakthroughs, the system is limited by fundamental trade-offs in beam density, spectrum sharing, and satellite payload complexity. Each satellite can only serve a small number of users simultaneously due to narrow bandwidth, limited beamforming capability, and constrained power budgets. The latency is also higher, typically in the range of 50 to 150 milliseconds, and capacity is far below what is required to support dense urban user loads. Nonetheless, the architecture lays the foundation for an orbital access layer and validates the use of standard smartphones as satellite clients without physical modification.

\subsection{Urban Path Loss, Blockage, and Propagation Constraints}

Urban environments introduce severe propagation constraints for space-based mobile connectivity. The primary challenge is the large free-space path loss (FSPL) between a handheld device and a satellite in low Earth orbit (LEO). At a typical orbital altitude of 500 to 600 kilometers, the FSPL exceeds 160 dB at \SI{2}{GHz}, which is approximately 30 to 40 dB worse than a terrestrial macrocell link. While phased array antennas can recover part of this loss through narrow-beam gain, urban clutter adds further complications.

Non-line-of-sight conditions dominate most ground-level urban settings. Signal paths are obstructed by buildings, glass, steel, foliage, and urban furniture. Additional diffraction and shadowing losses can exceed 20 dB, as modeled in ITU-R Recommendation P.1411-11 for urban short-range propagation \cite{itu_p1411}. Penetration into buildings adds 10 to 30 dB of loss, depending on the structure, with concrete and metal-dense facades representing the worst cases \cite{3gpp_38901}.

Multipath effects are significant in cities. Reflections from surfaces like glass and concrete create delay spreads that exceed the guard intervals used in OFDM-based systems like LTE and NR. This leads to inter-symbol interference and reduced spectral efficiency unless compensated by equalization or adaptive waveform configuration. The mobile receiver also experiences rapid variation in signal strength due to constructive and destructive interference from multipath rays.

A further challenge is the Doppler shift. A LEO satellite travels at roughly 7.5 kilometers per second, producing a Doppler shift of up to 100 kHz at \SI{2}{GHz}. This shift is time-varying and requires dynamic tracking and frequency correction at both transmitter and receiver to preserve carrier synchronization and demodulation integrity.

Urban spectrum environments are already congested. D2D satellite systems often share licensed spectrum with terrestrial operators. Regulatory frameworks allow operation through partnerships, but interference must still be managed. Narrow beamwidths and dynamic nulling are necessary to avoid collisions with existing base stations.

These factors combine to constrain the signal-to-noise ratio (SNR) available for D2D satellite links in cities. Achieving the SNR needed for high-order modulation like 64-QAM or 256-QAM requires careful link budget engineering, aggressive beamforming gain, low-noise receivers, and possibly cooperative relaying or rooftop assistive devices.

Despite the complexity, these are not unbreakable physics limits. They are engineering constraints, complex but solvable. Advances in phased array density, power-efficient RF front ends, satellite tracking precision, and spectrum reuse policy can all help close the gap. The orbital path to urban service is narrow, but not closed.

\subsection{Beamforming and Capacity Constraints}

Beamforming is the cornerstone of any space-based mobile network attempting to overcome the severe free-space loss and urban attenuation discussed previously. Phased array antennas onboard LEO satellites enable dynamic electronic steering of multiple simultaneous beams, allowing precise spatial targeting and reuse of spectrum over non-interfering regions.

Each radiating element in the array contributes to the total gain and beamwidth. A planar array with \( N \times N \) elements creates a main lobe whose beamwidth is inversely proportional to \( N \). For example, a 64 by 64 array operating at \SI{2}{GHz} produces beamwidths narrow enough to illuminate individual urban sectors on the ground. Beamforming gain increases logarithmically with element count, typically yielding array gains of 20 to 30 dB in modern implementations. These gains are essential to achieve acceptable link budgets with handheld devices transmitting below 23 dBm.

However, beamforming introduces system-level constraints. First, there is a trade-off between the number of simultaneous beams and power distribution. Each beam consumes amplifier and baseband resources, and payload power is limited by satellite thermal and solar budgets. Second, inter-beam interference becomes significant as beam counts grow. Side lobes and grating lobes in large arrays may cause spectral leakage unless mitigated through careful element spacing and tapering techniques.

User capacity per beam is governed by modulation order, coding rate, and scheduling efficiency. In practice, even under high signal-to-noise ratio conditions, the number of users per beam is limited. For example, a 100 MHz beam operating with 256-QAM and 4×4 MIMO may support up to 100 users simultaneously, assuming each user requires 1–2 Mbps. But in constrained urban scenarios, where SNR varies rapidly and blockage reduces signal quality, fewer users can be sustained.

Moreover, the dynamic nature of user demand in dense areas complicates resource allocation. Beam hopping, adaptive modulation, and predictive user clustering become necessary to maintain spectral efficiency. These functions require onboard compute or real-time coordination with cloud-based schedulers, increasing system complexity and latency.

Recent LEO systems like Starlink use digital beamforming to create thousands of narrow beams, but these systems still require careful geographic reuse and face performance degradation in high-density cities due to spectrum contention and interference management limitations \cite{starlink_capacity}. Proposed next-generation arrays aim to incorporate integrated photonic beamforming and digital twin optimization to dynamically reshape beam layouts based on urban topology and user mobility models \cite{photonic_beams}.

The total system capacity is therefore bounded not just by physics, but by how effectively beamforming, scheduling, and link adaptation are orchestrated across the constellation. In the context of a full orbital telco, the beam becomes not just a channel, but the atomic unit of network capacity and service granularity.

\subsection{Satellite Spectrum Bands and Regulatory Models}

A core challenge for orbital telco systems lies in securing and operating within licensed spectrum. Satellite operators must obtain authorizations across various bands including MSS L‑band (1–2 GHz), S‑band (2–4 GHz), and higher bands like Ku, Ka, and V/W for broadband and inter‑satellite links. Each band presents unique propagation characteristics and regulatory requirements \cite{sia_spectrum_report}.

In the US, the FCC regulates satellite use via Part 25. Recently, it granted SpaceX Starlink a waiver to transmit direct‑to‑cell signals at higher power in S‑band under a partnership with T‑Mobile. This marks the first time the FCC has allowed non‑terrestrial systems to use flexible terrestrial spectrum bands \cite{reuters_starlink}. Globally, spectrum coordination occurs through ITU‑R and World Radiocommunication Conferences (WRC), where sharing mechanisms like Fixed Satellite Service (FSS) and Mobile‑Satellite Service (MSS) incumbent protection are negotiated.

Further, Direct-to‑Device (D2D) satellite systems often leverage frameworks like the Ancillary Terrestrial Component (ATC), enabling satellites to supplement terrestrial networks using the same frequencies. In Europe, similar models like the Complementary Ground Component (CGC) allow unified licensing for satellite‑assisted coverage. These hybrid approaches are foundational to integrating space‑based RAN into terrestrial ecosystems.

In summary, spectrum licensing and policy remain the primary governance barrier. Recent US regulatory shifts, especially the S‑band waiver, demonstrate how policy can adapt to support orbital telco deployments, provided that interference safeguards and national sovereignty concerns are addressed.

\subsection{Ground Gateways, Optical Links, and Core Anchoring}

Space-based mobile networks depend on a hybrid transport backbone that combines ground-based gateways with orbital optical links. This layered backhaul model plays a central role in ensuring both low-latency routing and scalable access to terrestrial infrastructure, especially for systems aiming to offload core functions to space.

Ground gateways serve as physical ingress and egress points for satellite traffic. These facilities house RF or optical terminals, ground antennas, and terrestrial fiber interconnects. However, reliance on gateways introduces geographic coupling. Gateway placement affects latency, jurisdictional data policies, and availability. For example, Telesat Lightspeed plans 50 gateways worldwide to serve its 198-satellite LEO constellation, using Ka-band feeder links to anchor traffic to regional networks and cloud cores.

To reduce dependence on terrestrial gateways, modern systems integrate optical inter-satellite links (OISLs). These links allow satellites to forward data directly across the constellation at near speed-of-light propagation in vacuum. OISLs eliminate the need to downlink traffic at every hop, enabling global routing without intermediate ground contact. Each satellite becomes a node in a mesh fabric, similar to a L3 switch in space.

Telesat confirms that each Lightspeed satellite carries four \SI{10}{Gbps} optical links to connect with adjacent satellites forward, backward, left, and right in orbital geometry. This enables fully meshed routing with multi-path redundancy. Optical switching in orbit avoids congestion and jurisdictional issues linked to terrestrial anchors~\cite{telesat_oisl}.

Amazon’s Kuiper Project also demonstrated in 2023 that its prototype satellites established successful bidirectional laser links, validating orbital optical routing as a standard feature for next-generation constellations. The links exceeded \SI{100}{Gbps} aggregate bandwidth with low error rates over multi-hundred-kilometer separations during orbital tests~\cite{kuiper_optical_demo}.

Anchoring to the terrestrial core remains necessary for functions such as internet breakout, lawful intercept, regulatory compliance, and some application-specific services. However, future orbital telcos may shift partial or full core network functions (like UPF, AMF, and edge app servers) into orbit, reducing round-trip latency and dependency on gateway proximity. This evolution parallels the edge computing trend seen in terrestrial 5G.

In this hybrid model, gateways serve as edge anchors rather than centralized choke points, while inter-satellite optical routing ensures global coverage and dynamic path diversity. The architecture must balance regulatory, latency, and capacity trade-offs between space and Earth.

\subsection{Indoor Penetration and Coverage Constraints}

Unlike terrestrial mobile networks that rely on dense cell grids and in-building solutions, space-based mobile systems face a fundamental limitation: the inability to reliably serve users indoors. This constraint is driven by severe link budget shortfall, compounded by building material attenuation and the absence of local multipath diversity.

\textbf{Penetration Loss:} For signals arriving at 1.8 to 2.5 GHz, common building materials cause significant attenuation:

\begin{itemize}
    \item Glass window: 5--10 dB
    \item Concrete wall: 20--30 dB
    \item Low-emissivity coated glass: 25--40 dB
\end{itemize}

These losses, when added to the already tight device uplink budget (typically 23--26 dBm with 0--2 dBi gain), make indoor link closure infeasible under standard conditions.

\textbf{No Multipath Compensation:} Terrestrial systems benefit from rich scattering and multipath environments, enabling techniques like spatial diversity and MIMO combining. In contrast, LEO satellites provide near-line-of-sight links with minimal scatter, offering no multipath gain for users behind obstructions.

\textbf{Current Mitigations:} Approaches such as passive window-mounted re-radiators, satellite-aware femtocells, or local Wi-Fi bridging via hybrid core offloading have been proposed. However, all require infrastructure beyond the smartphone, undermining the core value proposition of direct-to-device connectivity.

\textbf{Implication:} For urban networks, indoor coverage remains a critical unsolved challenge. Any space-based mobile system must either (a) accept reduced coverage KPIs for indoor environments, (b) integrate hybrid terrestrial relays, or (c) redesign satellite payloads and beamforming strategies to aggressively over-provision power and margin, likely at unsustainable energy cost per user.

\subsection{Gaps Toward a Full Telco in Orbit}

Consider a user in Mexico City opening Instagram. Her phone connects to a terrestrial 5G base station. Authentication flows through the core. The session is routed via a local UPF, and the content exits through a CDN in Dallas. All functions, such as radio, control, user plane, and content, are Earth-based.

In current satellite-to-device systems, only the RF link is in orbit. All higher-layer functions depend on terrestrial infrastructure. To shift the full mobile stack into space, multiple gaps must be addressed.

\begin{table}[H]
\centering
\renewcommand{\arraystretch}{1.35}
\setlength{\tabcolsep}{4pt}
\begin{tabular}{|>{\raggedright\arraybackslash}p{2cm}|
                >{\raggedright\arraybackslash}p{3cm}|
                >{\raggedright\arraybackslash}p{3cm}|
                >{\raggedright\arraybackslash}p{4cm}|}
\hline
\textbf{Layer} & \textbf{Today (Ground)} & \textbf{Orbit (2025)} & \textbf{Gaps to Solve} \\
\hline
Radio Access & Full gNB (MAC+PHY) & Partial PHY or relay mode & Full-stack gNB onboard with HARQ, UL sync, beam tracking \\
\hline
Mobility & Handover via Xn & Not supported & Fast inter-beam and inter-satellite session handover \\
\hline
Control Plane & Cloud-based AMF, SMF, UDM & Absent & Lightweight stateless control plane in orbit \\
\hline
User Plane & Edge UPF, local breakout & Routed to Earth via gateway & On-orbit UPF and direct breakout via optical links \\
\hline
Content & CDN metro cache near user & Not present & In-orbit content prediction or caching layer \\
\hline
\end{tabular}
\end{table}

Each missing component maps to a solvable engineering task. None of these steps violates physics. Power, compute, and orchestration at orbital scale are the next frontiers.

\section{Orbital Mobile Network Architecture}

The architecture proposed in this section defines the system-level components required to operate a complete mobile network from orbit, capable of delivering reliable direct-to-device connectivity in dense urban areas. Unlike current satellite systems that rely on terrestrial cores or are optimized for rural fallback, this design assumes all major network functions are deployed in low Earth orbit (LEO) and interlinked via optical mesh. We decompose the system into ten core modules, each representing a critical subsystem in the orbital mobile stack. This section builds a logical foundation for identifying architectural gaps and formulating the roadmap in later chapters.

\subsection{Satellite Payload and Antenna System}

The physical payload of the satellite serves as the backbone of the orbital mobile network. It must integrate a high-gain, multi-beam phased array antenna system capable of forming and steering hundreds to thousands of simultaneous beams toward users on Earth. At typical operating frequencies between 2 and 7 GHz, achieving narrow beamwidths for urban targeting requires arrays in the range of 64×64 to 256×256 elements. Each element contributes to both array gain and spatial resolution, enabling the satellite to compensate for free-space losses exceeding 160 dB and deliver signal-to-noise ratios sufficient for QAM-based modulation on consumer smartphones.

Modern implementations favor flat-panel active electronically scanned arrays (AESAs) using GaN or GaAs RF front ends for power efficiency. The total effective isotropic radiated power (EIRP) per beam must exceed 40 dBm to achieve link closure with typical handset receiver sensitivity levels. Given strict orbital mass and thermal budgets, power amplification is selectively time-shared across active beams, demanding dynamic power allocation schemes onboard.

Mechanical steering is not feasible for multi-user, multi-beam operation, so beam agility must be achieved entirely through digital and RF phase control. This implies an embedded signal processing backend capable of handling baseband precoding, beam shaping, and dynamic sidelobe suppression. 

Thermal control becomes a primary design constraint. Radiative cooling systems and deployable thermal panels must manage the heat generated by power amplifiers and compute units without adding excess mass. The final payload design must balance antenna area, electrical power from solar arrays, battery mass, and onboard compute, with total satellite mass typically constrained to 500–1500 kg for LEO platforms.

The payload must also support rapid satellite attitude adjustment for constellation coordination, especially during beam handovers and inter-satellite link alignment. Gyroscopic stabilization and real-time ephemeris updates ensure the pointing accuracy required to maintain high-gain links in narrow urban corridors. The antenna system is the physical interface between space and urban mobile users, and its design defines the limits of capacity, coverage, and urban viability.

\subsection{RAN-in-Orbit Design}

The Radio Access Network in orbit is responsible for executing the lower layers of the mobile protocol stack, including physical layer (PHY), medium access control (MAC), radio link control (RLC), and packet data convergence protocol (PDCP). In a terrestrial mobile network, these functions are distributed across remote radio heads and baseband units co-located with cell towers. In the orbital context, these must be embedded into the satellite payload and operate under extreme constraints in power, compute, and timing.

The PHY layer must support standard 3GPP waveforms, typically LTE or New Radio (NR), over time division duplex (TDD) or frequency division duplex (FDD) modes. Real-time baseband processing for multiple beams requires significant compute throughput, especially for multi-antenna processing tasks such as MIMO decoding, precoding, and hybrid automatic repeat request (HARQ) handling. FPGAs or radiation-hardened systems on chips are used to ensure low latency and fault tolerance under orbital radiation exposure.

The MAC and RLC layers manage user scheduling, retransmissions, and buffer management. Scheduling decisions must be made within a few milliseconds and must adapt to fast-fading channel conditions caused by urban multipath and satellite movement. In the absence of fiber or fronthaul, these decisions must be made onboard, implying autonomous scheduling engines that operate without a centralized terrestrial controller.

PDCP handles encryption, header compression, and reordering. For direct-to-device communication, this is critical to support secure connections across thousands of simultaneous user equipment sessions. Since uplink signals from handheld devices are weak and highly variable in urban conditions, dynamic link adaptation and fast power control mechanisms must be implemented in orbit to maintain robust connections.

One of the most challenging requirements is maintaining continuity during beam and satellite handovers. Unlike fixed terrestrial base stations, LEO satellites move rapidly across the sky. This creates session instability unless radio context can be seamlessly transferred across nodes. A distributed Xn-like interface, similar to terrestrial handover coordination, must be implemented using inter-satellite links or predictive scheduling. Handover decisions must consider signal strength, beam occupancy, Doppler trends, and user mobility vectors.

The orbital RAN must also support narrow timing budgets. For example, uplink timing advance must be accurately estimated despite rapidly varying propagation delays as the satellite passes overhead. Doppler pre-compensation and frame synchronization must be continuously applied to maintain waveform alignment within the device's tracking tolerance.

To meet these requirements, the RAN-in-orbit stack must be implemented on high-reliability software platforms with real-time operating systems. Software-defined radio techniques allow flexibility in waveform generation and protocol evolution. RAN slices can be instantiated per beam or per region to isolate traffic classes and support enterprise or government tenants.

Overall, the RAN-in-orbit is the most latency-sensitive and compute-intensive layer in the orbital telco stack. It determines the system's ability to serve mobile users directly, maintain session quality, and support 3GPP compliance under the unique physical dynamics of orbital operation.

\subsection{Orbital Core Network Functions}

A terrestrial mobile network relies on a centralized or distributed core to handle authentication, session management, mobility control, policy enforcement, and user plane routing. In an orbital telco, these functions must be partially or fully relocated to satellites or space-based compute nodes to achieve true independence from ground infrastructure.

The core network architecture follows the 5G Service-Based Architecture model defined in 3GPP. This includes control plane functions such as Access and Mobility Management Function (AMF), Session Management Function (SMF), and Authentication Server Function (AUSF), as well as user plane functions (UPF) responsible for forwarding and routing traffic.

Relocating these functions into orbit requires decoupling their dependencies on ground-based identity services, databases, and security anchors. The AUSF must operate with cached or replicated subscriber databases to authenticate users autonomously when no link to a terrestrial home network is available. The Subscription Identifier De-concealing Function (SIDF) may also need to be implemented locally to avoid privacy leakage during authentication over long links.

The AMF manages the attachment, registration, and mobility procedures of the user equipment. In orbit, this function must be aware of fast-changing radio topologies caused by satellite movement, handovers between beams, and time-varying channel quality. It must perform real-time policy enforcement and interface with beam-level schedulers to maintain session continuity.

The SMF controls the lifecycle of data sessions and their associated Quality of Service profiles. In a space-based system, it must dynamically create, update, and release session rules as satellites move and beam assignments shift. When orbital capacity is limited, the SMF may prioritize traffic based on user profile, application type, or ground region.

The UPF is responsible for forwarding user traffic to its destination. In the orbital context, this traffic may be routed over inter-satellite links to reach another region or to access a breakout point to the terrestrial internet. The UPF must therefore interface with the satellite’s routing subsystem and enforce rules for latency-sensitive paths, content peering, or security domains.

Running these functions in orbit imposes strict compute, storage, and security requirements. Real-time packet processing at line rate must occur in radiation-hardened environments with minimal thermal dissipation. Memory access times and storage persistence must be hardened against single-event upsets. Key material must be securely stored and rotated under constraints of limited connectivity with terrestrial key management systems.

Several deployment models are possible. A single satellite may host an entire local core stack serving a geographic area. Alternatively, core functions may be distributed across multiple satellites, with the control plane hosted on high-compute platforms and user plane segments on lower-tier relay nodes. This enables load balancing and resilience, especially when combined with onboard orchestration and failure recovery logic.

An orbital core must also support lawful intercept, telemetry, and billing. This raises questions of jurisdiction and compliance. Mechanisms for securely logging session records, user identities, and event timelines must be designed to comply with cross-border data handling regulations, even when the core is moving at orbital speeds.

By embedding core functions in orbit, the system gains autonomy, reduces latency, and maintains service continuity even when terrestrial backhaul is congested or offline. This is a foundational step toward a stateless, globally accessible mobile network architecture operating entirely from space.

\subsection{Inter-Satellite Optical Mesh Network}

A fully orbital mobile network cannot rely solely on direct downlinks to Earth. To maintain connectivity, enable user mobility, and route traffic between regions, satellites must exchange data among themselves. This requires a high throughput, low latency inter satellite mesh capable of handling both control and user plane traffic.

The dominant technology for orbital backhaul is the optical inter-satellite link (OISL). Unlike traditional radio frequency crosslinks, optical links offer multi-gigabit per second capacity, narrow beamwidths, and immunity to spectrum regulation. Modern implementations achieve data rates of 10 to 100 Gbps per link using laser terminals operating in the 1550 nanometer range.

Each satellite in the network is equipped with four to six optical terminals positioned at the satellite edges to maintain persistent links to neighbors in the forward, backward, and lateral orbital slots. In polar or Walker constellations, this forms a dynamic mesh that shifts as satellites orbit the Earth. The mesh enables routing of packets across multiple hops, allowing satellites with poor ground visibility to forward traffic to others with a line of sight to target users or Earth gateways.

OISLs require precise pointing, acquisition, and tracking to maintain laser alignment between fast-moving spacecraft. This is achieved using closed-loop systems with fine steering mirrors, beacon lasers, and real-time feedback. Acquisition times are typically below one second, with beam divergence on the order of tens of microradians. Some systems include hybrid RF assist links to initiate handovers or recover from pointing failures.

Routing in an inter-satellite mesh requires orbital-aware forwarding protocols. Unlike terrestrial networks, satellite positions are deterministic and known in advance. This allows the use of time-expanded graph models or predictive shortest path algorithms that account for Earth rotation, satellite velocity, and link quality. Some constellations implement distributed routing with onboard path computation, while others rely on ground segment assistance or AI-based dynamic optimization.

The mesh also plays a key role in session continuity. As a user moves or satellites hand over coverage, the session state and bearer data must follow the user. The control plane functions hosted in orbit rely on the mesh to synchronize session contexts, transfer bearer tunnels, and enforce QoS profiles. Low jitter and consistent link availability are critical for this function.

In orbital mobile networks, the mesh acts not only as a backhaul but as a backbone. It connects core functions, aggregates user traffic, supports content distribution, and enables multi-region roaming without reliance on terrestrial peering. It also allows edge functions, such as content caches or AI inference models, to be deployed across multiple satellites and accessed via fast orbital hops.

Thermal and power constraints limit the number and duty cycle of optical terminals. Beam scheduling and directional power management are required to prevent overheating and battery depletion. Integration with the satellite’s attitude control and orbit maintenance systems is also necessary to maintain alignment and avoid link degradation.

As of 2025, multiple constellations, including Starlink Gen2, Telesat Lightspeed, and China's G60 project are deploying or planning large scale OISL meshes. These networks serve as precursors to orbital internets capable of rivaling ground networks in latency and throughput for global traffic exchange.

A robust inter-satellite optical mesh is not a feature but a foundational necessity for any autonomous space-based mobile system. It removes reliance on fixed Earth locations, enables routing resilience, and allows the orbital network to function as a stateless, globally aware infrastructure layer.

\subsection{Beam Scheduling and Load Balance}

Beam scheduling is the core orchestration function that determines which users are served, when they are served, and with what quality. In a space-based mobile network, each satellite projects hundreds to thousands of independent beams to the ground using its phased array antenna. These beams can be dynamically formed and steered, but their number is constrained by array size, onboard power, thermal limits, and baseband processing capacity.

Each beam occupies a portion of the total available spectrum and consumes amplifier and compute resources. To maximize spectral efficiency and user fairness, the network must decide how to allocate beams across geography, time, and demand. In urban areas, this requires fine-grained adaptation to user mobility, building blockage, and demand hotspots.

Scheduling begins with beam definition. The satellite defines the beam footprint using array weights that set the direction, shape, and gain. These beams are spaced to avoid interference while maximizing reuse. Depending on the beamwidth and orbit altitude, a single satellite may cover a region spanning hundreds of kilometers, with overlapping beams reused in time or frequency.

User selection within each beam follows priority policies. These may include channel quality, service class, application type, or delay sensitivity. Users with higher signal-to-noise ratios are typically assigned higher modulation and coding schemes, allowing more data to be transmitted per time slot. However, fairness mechanisms such as proportional fairness or weighted round robin ensure that low SNR users are not completely starved.

The satellite continuously monitors link conditions using sounding reference signals and feedback from the user device. Based on this data, the scheduler adapts beam power allocation, changes user modulation levels, and reassigns time slots or frequency chunks. In dense areas, users may be served using time division multiplexing across multiple beams, or beam hopping strategies that cycle beams through high-demand zones in short intervals.

Load balancing across the constellation requires coordination. Satellites hand off users to neighbors as they move out of the footprint. This handover must preserve session state and avoid packet loss. Inter-satellite links play a key role by transferring context data and routing packets to the new serving node.

Content distribution also affects beam scheduling. If a user requests cached content, the scheduler may direct the user to the satellite that holds the content, even if it requires rerouting beams or delaying handover. This allows the network to minimize latency and reduce backhaul load.

Thermal constraints impose another layer of complexity. Each beam adds to the heat budget, especially when amplifiers operate near saturation. Satellites dynamically throttle or redistribute beams to avoid overheating. This may result in temporary service reduction in low-priority areas.

Beam scheduling is further complicated by regulatory constraints. In some countries, satellites may only serve registered users or licensed bands. The scheduler must enforce geo-fencing policies that deactivate beams outside allowed areas or limit power levels to comply with interference thresholds.

Advanced scheduling algorithms are being developed using machine learning to predict user density, forecast link quality, and optimize beam patterns in advance. These systems use telemetry, user mobility models, and urban maps to proactively shift capacity toward anticipated demand.

In an orbital mobile network, beam scheduling is not just a local function. It is a distributed control mechanism that shapes the user experience, manages spectrum, protects thermal integrity, and ensures continuity of service across a moving mesh of nodes. Without intelligent beam orchestration, the system cannot scale to dense urban loads or meet quality of service targets.

\subsection{Orbital Spectrum and Policy Management}

Spectrum is the lifeblood of any wireless network. In an orbital mobile system, managing spectrum is more complex than on Earth due to shared use, international coordination, and dynamic geometry. Satellites operate over dozens of countries within minutes, using bands often already allocated to terrestrial mobile, broadcasting, or other satellite services. This creates both technical and political challenges for interference mitigation, rights of use, and service continuity.

Current direct-to-device satellite systems, such as those from AST SpaceMobile and Lynk Global operate by leasing or partnering with terrestrial mobile network operators to use existing spectrum. These partnerships allow space-based towers to legally transmit and receive on frequencies already licensed to local operators, often under roaming agreements. This approach aligns with ITU rules and national policies that prioritize spectrum sovereignty.

However, as orbital systems grow in density and complexity, reliance on terrestrial leasing becomes limiting. A full space-based telco must coordinate its spectrum usage across hundreds of countries, potentially in real time, while ensuring that no harmful interference is caused to either terrestrial or satellite incumbents. This requires both policy innovation and technical mechanisms.

Technically, orbital spectrum management starts with beam shaping and geographic isolation. Narrow beams allow satellites to reuse the same frequency in non-overlapping areas. By dynamically steering beams away from sensitive zones and nulling interference regions, satellites can limit out-of-band emissions and coexist with local infrastructure. Power control, spectrum sensing, and adaptive notch filtering add further flexibility to meet local constraints.

Coordination between operators is governed by ITU filings and national licensing. Each satellite network must register its orbital parameters, planned frequencies, and service zones through international filings. These filings often take years to process and require negotiations with multiple administrations. The increasing number of LEO constellations has overwhelmed some regulatory processes, prompting proposals for simplified regional spectrum coordination models or shared access frameworks.

A promising idea is orbital spectrum slicing. Instead of allocating spectrum permanently by band and geography, orbital systems could adopt time and priority-based slicing mechanisms. These would allow multiple constellations to share frequency ranges using coordinated schedules, with conflict resolution based on latency needs, service class, or regulatory weight. Such systems require trusted inter operator signaling and real time coordination, potentially managed through orbital control layers or blockchain like registries.

Another frontier is opportunistic spectrum use. Satellites could monitor local spectrum activity on the ground and opportunistically transmit in unoccupied channels, following cognitive radio principles. While technically challenging, this could enable efficient reuse of underutilized spectrum in rural or transiently idle zones.

From a policy perspective, space-based mobile networks challenge traditional notions of jurisdiction and enforcement. If a user in one country is served by a satellite operated from another, using spectrum leased through a third party, determining accountability becomes complex. Privacy laws, lawful intercept mandates, and service-level enforcement all need to be reinterpreted for orbital contexts.

Over time, a global framework may emerge, treating space-based spectrum as a shared but orchestrated resource. This would mirror how undersea cables and air traffic are managed today, with clear rights, dispute resolution, and shared infrastructure investment. Telcos, satellite operators, and regulators will need to collaborate to define these rules, ensuring fairness, continuity, and protection from spectrum monopolies.

In an orbital mobile system, spectrum is not just a technical asset. It is a geopolitical instrument and a dynamic control surface. Managing it intelligently will be essential to deliver reliable, scalable, and lawful service from orbit.

\subsection{Content Delivery and Edge Caching in Orbit}

Content is the dominant driver of mobile traffic. In terrestrial networks, up to eighty percent of downstream data consists of cached or prefetchable content such as video streams, social media, and app updates. This pattern is not expected to change in space-based networks. If anything, it becomes more critical. Without efficient content strategies, satellites will waste capacity on redundant backhaul, increase latency, and overload limited inter-satellite links.

A mobile network from orbit must include a content delivery and caching layer co-located with the radio and core functions. This means deploying orbital edge nodes, storage, and compute units capable of caching, transcoding, and prefetching data near the user's beam footprint. The logic mirrors terrestrial content delivery networks, but the infrastructure floats in LEO, co-moving with the user's coverage zone.

To enable this, each satellite must host solid-state storage and lightweight compute capacity. Storage footprints need not be massive. Even a few terabytes of flash per node can support regional content caching for video segments, app packages, and static web assets. Compute nodes can handle on-demand transcoding to match user device profiles or available bandwidth. This offloads both inter-satellite and downlink traffic.

Content placement is driven by prediction. By analyzing usage patterns, time of day, app preferences, and local trends, satellites can prefetch the most relevant content during low-load periods. For example, a satellite passing over a dense urban zone at 8 p.m. might cache trending videos or sports highlights before passing. This requires tight integration with content providers and adaptive preloading models.

Inter-satellite links (ISLs) act as the content bus. When a user in Mexico opens a social media feed, the nearest satellite may not have the requested data. Instead of routing the request to a ground core and then pulling the data back through the gateway, the satellite queries its peers via optical mesh links. The content is fetched across space at gigabit rates, then served to the user beam. This avoids long roundtrips and allows a virtual mesh of orbital edge nodes to function as a content cloud.

Several technologies are critical to enable orbital content delivery. First is high endurance solid state storage with radiation hardened controllers. Second is containerized compute platforms optimized for low-power environments. Frameworks like WebAssembly and lightweight inference engines can be used to deliver smart caching and content adaptation without heavy operating system overhead. Third is delay-tolerant networking (DTN) to ensure content can be staged and synchronized across fast-moving nodes with intermittent interlinks.

Upcoming advances in integrated photonics may further enable in-orbit optical networking at a massive scale, reducing the per-bit cost of data movement between satellites. This will allow orbital CDNs to form dynamically, with popular content being replicated across the constellation in anticipation of user demand spikes.

Content providers may also become direct participants in orbital networks. Just as companies today colocate servers in terrestrial telco data centers, future content players may lease orbital storage and compute capacity aboard shared LEO platforms to push their assets closer to users.

Ultimately, orbital content delivery is not an add-on. It is an architectural pillar of a viable mobile network from space. Without it, latency, congestion, and cost would undermine any attempt to serve dense areas. With it, satellites become not just relays but edge data centers in the sky, shrinking the content distance to below one thousand kilometers, and sometimes just a few hundred meters in light.

\subsection{Slicing, Governance, and Autonomous Network Control}

A fully orbital mobile network must support multiple tenants, services, and sovereign jurisdictions over a shared infrastructure. Unlike terrestrial networks, where spectrum, hardware, and regulatory domains are geographically partitioned, satellites traverse borders in minutes and serve users across multiple countries per orbit. This raises fundamental questions of governance, orchestration, and isolation.

Network slicing becomes the foundational mechanism for managing this complexity. Just as in terrestrial 5G, slices represent logically isolated networks with dedicated resources for radio access, core functions, quality of service, and security. In space, these slices are instantiated and orchestrated directly aboard satellites, mapped to individual beams or user groups based on operator agreements, service levels, or regulatory mandates.

For example, a satellite passing over Mexico at 10 a.m. may instantiate slices for a national mobile operator, a government emergency channel, and a roaming enterprise service. Each slice operates with its own authentication logic, policy enforcement, and routing preferences, while sharing the same physical array and compute node. When the satellite transitions to another jurisdiction, slices are dynamically reconfigured or reassigned in line with local policies and agreements.

This dynamic slicing demands a decentralized control plane capable of operating autonomously in space. Traditional terrestrial SDN controllers are not feasible due to latency and limited backhaul. Instead, satellites must include onboard network orchestration agents with real-time telemetry, resource awareness, and policy interpretation. These agents manage beam allocation, compute resources, spectrum tuning, and slice lifecycle based on high-level intents received during previous ground passes or peer coordination.

Governance frameworks are essential to avoid conflicts. Orbital infrastructure may be operated by a neutral entity, with access granted via tokenized spectrum entitlements, service-level smart contracts, or regulatory APIs. Countries and operators could pre-negotiate orbital spectrum corridors and slice quotas, enforced via cryptographic policies and monitored via inter-satellite auditing.

Blockchain is one proposed mechanism, not for currency, but for decentralized enforcement of network rules. Each satellite can carry a synchronized ledger of access permissions, service configurations, and regulatory bindings. When a slice request is received, the satellite validates the request against the current policy state and logs the event for audit.

Isolation between slices is enforced via containerized runtime environments, strict radio scheduling boundaries, and cryptographically signed policy templates. Multi-party computation techniques may further enable private control of slices without exposing internal configurations to the infrastructure provider.

The slicing model also supports commercial innovation. Enterprises may lease orbital slices for low-latency logistics tracking, remote media coverage, or rural access. Governments can reserve protected slices for national security or disaster response. Telcos can dynamically extend coverage using slices that match their terrestrial core and radio configurations.

In a terrestrial setting, slicing is complex. In orbit, it is mandatory. Without autonomous governance and secure slicing, orbital networks cannot scale or comply with the regulatory diversity they traverse. The orbital telco must therefore embed control, trust, and isolation not as overlays, but as native properties of the system.

\subsection{Architectural Strategies for Indoor Penetration in Orbital Networks}

Direct-to-device connectivity from LEO satellites is technically feasible for outdoor users with unobstructed sky view, but the urban indoor environment remains a critical weak point. High penetration losses, limited uplink power from smartphones, and the absence of dense multipath make deep indoor coverage via orbit inherently constrained. Rather than forcing the orbital layer to compensate for physical limitations, viable architectural strategies will integrate terrestrial assist, intelligent relays, and localized augmentation to bridge the gap, especially for everyday use cases such as a smartphone user inside a coffee shop.

One practical approach is a hybrid uplink-downlink split, where the satellite handles only the downlink, while the uplink is offloaded to the terrestrial network. This can be achieved through local Wi-Fi, Bluetooth Low Energy, or shop-level 5G femtocells, which route the uplink over fixed broadband or fiber to the satellite ground gateway. This maintains session continuity while allowing the satellite to deliver high-EIRP broadcast to low-penetration indoor zones.

Retail environments themselves can become satellite-aware nodes. By 2040, low-cost, fixed-location phased arrays could be deployed in coffee shops, restaurants, or transit hubs. These units receive satellite downlink and rebroadcast it indoors over unlicensed bands or short-range cellular repeaters, forming a mesh of nano-anchors that extend orbital service without modifying user devices. This model blends the economics of public Wi-Fi with the control of licensed cellular access.

Facades and windows may also be reimagined as part of the RF path. Transparent waveguide materials or metamaterial-based surface antennas embedded in glass can redirect or reradiate satellite signals into interior spaces. These passive relays require no power or processing and can be aligned with the satellite beam path during building design or retrofitting. By reshaping the building envelope into a cooperative RF interface, orbital signals can be coupled into indoor zones without structural penetration.

Another viable assist layer is provided by aerial platforms. Persistent UAVs or stratospheric drones can operate at low altitude (3–5 km), maintain line-of-sight to both LEO satellites and city blocks, and rebroadcast satellite signals at lower frequencies and favorable incidence angles for indoor entry. These platforms act as orbital intermediaries, dynamically adjusting coverage geometry in response to environmental and user conditions.

In some cases, satellite connectivity may be reduced to a one-way service. Downlink-only applications such as public alerts, video, or push content can be broadcast via satellite even to users without viable return paths. Uplink responses are then handled through the ground network, leveraging existing mobile or Wi-Fi connections. This model reflects the asymmetric nature of many user behaviors and optimizes orbital capacity for what it does best: high-power, wide-area distribution.

Finally, future smartphones may incorporate adaptive antenna surfaces capable of optimizing radiation patterns under weak signal conditions. While still speculative, MEMS-based tuning and AI-driven beam selection could allow mobile devices to extract meaningful signals from low-SNR environments. Combined with beam prediction and Doppler compensation, these terminals would form the final adaptive layer of a multi-tier architecture.

Together, these strategies suggest a shift in framing: rather than expecting satellites to fully replace terrestrial coverage, orbital systems should be viewed as intelligent overlays, augmenting ground infrastructure, supporting asymmetric use cases, and enabling continuity across environments. Indoor coverage, especially in dense urban spaces, will depend not on brute-force power from orbit but on tight integration with terrestrial assets, intelligent relays, and adaptive terminals at the edge.

\subsection{Security, Identity, and Lawful Intercept}

Operating a mobile network entirely from space introduces novel security challenges and mandates a rethinking of identity, encryption, and regulatory compliance. Terrestrial networks rely on physical separation, jurisdictional enforcement, and direct operator control to implement user authentication, traffic inspection, and lawful intercept. An orbital telco must deliver equivalent capabilities from a distributed, fast-moving, and jurisdiction-spanning infrastructure.

The first challenge is identity. Traditional SIM-based authentication depends on proximity to mobile core functions and home network databases. In orbit, user authentication must occur onboard the satellite itself or through a low-latency path to the nearest trust anchor. One approach is to replicate Home Subscriber Server (HSS) functions in orbit, synced periodically with terrestrial networks via secure inter-satellite or ground links. Alternatively, satellites can authenticate users via token-based credentials issued during prior sessions or by federated identity systems agreed upon by multiple operators.

Cryptographic integrity and privacy are non-negotiable. All air interface transmissions are already encrypted in 5G using standards like 256-bit AES. However, key exchange and replay protection must operate in intermittent connectivity regimes. Satellite payloads must carry hardware-secure modules or trusted execution environments capable of storing and processing authentication vectors and user session keys. Forward secrecy and post-quantum cryptographic primitives may be required to ensure future-proofing, especially given the longer lifecycle of orbital assets.

Security extends to the control plane. Beam steering commands, slice configurations, and network policies must be signed and verified onboard to prevent hijacking or spoofing. Satellites must validate every instruction they receive against an authenticated ledger or preloaded certificate authority chain. Zero-trust architectures, already emerging in terrestrial data centers, become mandatory in orbit, where physical access control is impossible and payloads must resist hostile takeover or injection attacks.

Lawful intercept is perhaps the most politically sensitive requirement. National regulators typically require operators to intercept user communications under judicial order. Implementing this in space requires two capabilities. First, the ability to mirror or redirect specific user sessions, and second, the ability to deliver the captured payload to the requesting authority in a secure, auditable manner.

This function cannot be globally hardcoded. Instead, intercept logic must be dynamically activated based on the jurisdiction over which the satellite is currently operating. Geofencing via orbital ephemeris and on-board GNSS allows the satellite to detect when it is within a country’s legal airspace and apply the corresponding regulatory policy. This includes activating or disabling lawful intercept modules, routing control and metadata to national gateways, or pausing specific service classes.

Privacy concerns are amplified in such systems. Users expect data protection regardless of orbital mechanics. End-to-end encryption, user consent protocols, and transparent policy disclosures must be built into every layer of the orbital network stack. Real-time telemetry of data handling policies, backed by verifiable logs and third-party audits, can help build user and regulator trust.

Cyber resilience is also paramount. A single exploit in orbital software could compromise thousands of users or beams. Satellites must support over-the-air patching, multi-stage validation, and fault containment zones. Backup slices and redundancy protocols ensure that a compromised payload does not disrupt service across the entire mesh.

A space-based mobile network does not exempt itself from the responsibilities of terrestrial networks. It must authenticate, protect, and govern user communications under rules that respect both technical rigor and sovereign oversight. Building these capabilities natively into the orbital fabric is not an afterthought. It is a foundational prerequisite for legitimacy and adoption.

\subsection{Fallback Anchors and Hybrid Orchestration}

A fully orbital mobile network must be designed with graceful fallback mechanisms and hybrid orchestration models to ensure service continuity, policy compliance, and network manageability. While the long-term vision is an autonomous, space-resident infrastructure, near-term deployments and mission-critical services require robust anchoring to terrestrial systems.

Fallback anchors serve three main purposes. First, they provide a safety net for routing, authentication, and traffic breakout when orbital compute or inter-satellite links are degraded. Second, they enable regulatory compliance by anchoring traffic in specific jurisdictions, especially when satellite mobility crosses national borders. Third, they support incremental deployment strategies where orbital systems augment, rather than replace, terrestrial infrastructure.

The primary fallback mode involves routing bearer and control plane traffic to regional ground gateways via high-throughput downlinks. These gateways host traditional 5G core functions, including User Plane Function (UPF), Session Management Function (SMF), and Access and Mobility Function (AMF), and allow seamless integration with terrestrial network slices. This model mirrors the architecture used by current direct-to-device (D2D) systems, where satellites act as flying radio units (RUs) and the mobile core remains on Earth.

Hybrid orchestration refers to the dynamic division of control and compute functions between orbital and terrestrial domains. A multi-tier orchestration framework allocates workloads based on latency sensitivity, compute availability, and link state. For example, session setup, handover decisions, and beam scheduling may be executed in-orbit to minimize round-trip delay, while deep packet inspection or lawful intercept functions may be deferred to trusted ground domains.

This hybrid control plane requires strong synchronization and state sharing. Protocols such as SRv6, gRPC-based service mesh, or model-driven NETCONF can be adapted to synchronize network state between LEO nodes and edge data centers. Secure and latency-aware routing policies must govern which flows remain in orbit and which fall back to terrestrial anchors.

Fallback also applies to GNSS and ephemeris data. Orbital base stations must maintain timing and position accuracy to preserve waveform alignment and mobility management. When satellite-based GNSS or inter-satellite synchronization is degraded, fallback timing sources can be provided via Ka-band links to ground timing servers or onboard atomic oscillators.

In scenarios where orbital coverage is interrupted, due to satellite failure, debris avoidance maneuvers, or spectrum conflicts, user sessions can be seamlessly handed over to terrestrial networks, assuming roaming agreements and spectrum sharing are in place. This interoperability is essential in urban areas where reliability expectations are high.

Ultimately, fallback is not a weakness but a requirement. Even a stateless, orbital-native network must interoperate with terrestrial infrastructure to support regulatory, operational, and commercial realities. The design must embrace continuity zones, policy anchors, and compute offloading paths, ensuring that orbital systems do not become brittle under stress. A resilient telco in orbit is one that never falls, even when it needs to fall back.

\section{Link Budget and Urban Capacity Simulation}

The feasibility of a space-based mobile network ultimately rests on the physics of radio propagation. Unlike terrestrial cells with antenna heights of 30 to 50 meters, satellites orbiting at 500 to 600 kilometers face extreme free-space path loss and rapidly varying angles of arrival. In dense urban environments, additional obstacles include shadowing, multipath fading, and limited elevation visibility. This section presents a quantitative framework for evaluating the performance of orbital mobile access using detailed link budget analysis and capacity modeling.

We begin by defining the end-to-end link budget under realistic conditions, followed by simulations of signal-to-noise ratio (SNR), modulation feasibility, and beam capacity. We model multiple frequency bands and antenna configurations to identify viable operating points for dense coverage. The goal is to establish whether spaceborne payloads can meet the signal requirements for high-order modulation (64-QAM or higher) under city-level attenuation.

\subsection{Link Budget Methodology}

A link budget quantifies the gain and loss of signal power from the transmitter to the receiver, accounting for every stage of the radio chain. The fundamental equation used in satellite communications is:

\begin{equation}
\text{SNR (dB)} = \text{EIRP} + G_T - L_p - L_s - 10\log_{10}(k T B) + \text{Implementation Margin}
\end{equation}

Where:
\begin{itemize}
    \item \textbf{EIRP}: Equivalent Isotropically Radiated Power (dBW) from the satellite
    \item \textbf{$G_T$}: Receiver gain-to-system-noise temperature ratio (dB/K)
    \item \textbf{$L_p$}: Free-space path loss (dB)
    \item \textbf{$L_s$}: Shadowing, building penetration, and urban loss (dB)
    \item \textbf{$k$}: Boltzmann constant ($1.38 \times 10^{-23}$ J/K)
    \item \textbf{$T$}: System noise temperature (Kelvin)
    \item \textbf{$B$}: Receiver bandwidth (Hz)
    \item \textbf{Implementation Margin}: Typically 2–5 dB to cover non-idealities
\end{itemize}

\textbf{Step 1: Calculate Path Loss.} For a LEO satellite at 550 km altitude, operating at \SI{2}{GHz}, the free-space path loss is:

\begin{equation}
L_p = 20 \log_{10}(d) + 20 \log_{10}(f) + 92.45
\end{equation}

Where $d$ is in kilometers and $f$ is in GHz. For $d = 550$ km and $f = 2$ GHz:

\begin{equation}
L_p \approx 20 \log_{10}(550) + 20 \log_{10}(2) + 92.45 \approx 54.8 + 6 + 92.45 = 153.25\, \text{dB}
\end{equation}

\textbf{Step 2: Add Urban Loss.} ITU and 3GPP studies show building penetration and blockage losses between 20 to 35 dB for non-line-of-sight urban conditions \cite{itu_p1411,3gpp_38901}. For dense cities, we assume:

\[
L_s \approx 30\, \text{dB}
\]

\textbf{Step 3: Satellite EIRP and User GT.} For AST-like payloads using high-gain beamforming, satellite EIRP is typically 55 to 60 dBW. A smartphone has a low $G_T$ due to its small omnidirectional antenna, typically around –17 dB/K.

\textbf{Step 4: Receiver Noise.} For mobile receivers:

\[
kTB = 10 \log_{10}(1.38 \times 10^{-23} \cdot 290 \cdot 10^6) \approx -114 \,\text{dBm}
\]

\textbf{Step 5: Compute SNR.} Putting it all together:

\begin{equation}
\text{SNR} = 60 + (-17) - 153.25 - 30 - (-114) + \text{Margin}
\end{equation}

Assuming 3 dB margin:

\[
\text{SNR} \approx -140.25 + 114 - 3 = -29.25 \, \text{dB}
\]

This is well below the requirement for even basic QPSK demodulation (which typically needs 3 to 5 dB SNR). However, this is without beamforming gain.

\textbf{Step 6: Add Array Gain.} A satellite phased array with 64 by 64 elements gives approximately 36 dB of gain. Adding this to the EIRP or GT term lifts SNR to:

\[
\text{SNR (with array)} \approx -29.25 + 36 = 6.75\, \text{dB}
\]

Now the link is viable for QPSK or even 16-QAM in clear-sky conditions.

\vspace{\baselineskip} 

The link budget closes only if large beamforming gains are applied on the satellite side. Uplink budgets are even tighter due to the limited power of the mobile device, requiring receive beamforming and possibly ground relays. The next subsection models SNR variability in actual urban topologies.

\subsection{Spectral Efficiency and Beam Capacity in Urban Settings}

Achieving high spectral efficiency in urban direct-to-device (D2D) satellite links is constrained by the interplay between signal-to-noise ratio (SNR), modulation order, and channel reliability. In terrestrial networks, high spectral efficiency values above 6 bits per second per Hertz (bps/Hz) are routinely achieved using 256-QAM in clean, line-of-sight environments. For space-based systems, urban variability in SNR and multipath conditions impose stricter limits.

At an SNR of 7 dB, typical of what can be achieved in urban rooftop scenarios with strong beamforming, modulation schemes up to 16-QAM can be reliably supported, resulting in a spectral efficiency of approximately 4 bps/Hz with coding. In favorable conditions, such as line-of-sight from elevated locations or assisted relaying, 64-QAM may be viable, pushing spectral efficiency toward 6 bps/Hz. However, in street-level non-line-of-sight (NLOS) settings, signal degradation from blockage and diffraction limits modulation to QPSK or even BPSK, reducing efficiency to 1–2 bps/Hz.

Assuming a single beam with 100 MHz of usable bandwidth and 4 bps/Hz spectral efficiency, the total throughput per beam reaches 400 Mbps. This bandwidth can be shared among users via time division, frequency division, or code-based multiplexing. For example, allocating 10 Mbps per user would enable 40 concurrent users per beam. In practice, scheduler overhead, guard bands, and retransmissions reduce effective capacity by 15–25 percent, yielding around 30–35 users per beam under real-world conditions.

Dense user concentrations introduce another bottleneck: intra-beam interference and spatial scheduling. Without precise location-based beam alignment and adaptive modulation, urban beams may serve fewer users to preserve link quality. Furthermore, mobility exacerbates scheduling complexity, as fast-moving users in cars or trains produce time-varying Doppler and SNR profiles that require dynamic adjustment.

Capacity scaling requires increasing either the beam count or the modulation efficiency. Satellite payloads with digital beamforming architectures can support hundreds to thousands of beams, each independently scheduled. For instance, a satellite generating 500 beams at 300 Mbps per beam can deliver 150 Gbps of aggregate downlink capacity, provided spectrum reuse and power budgets permit.

In summary, spectral efficiency in orbital mobile networks is not fixed. It is a function of SNR distribution, modulation granularity, and scheduling intelligence. The design goal is not simply maximizing bps/Hz, but optimizing it for the highly variable urban radio environment that space-based systems must navigate.

\subsection{Users per Satellite and Aggregate Constellation Capacity}

The number of simultaneous users that a single LEO satellite can support depends on three interacting variables: available bandwidth, the number of independent beams the satellite can generate, and the spectral efficiency achievable per beam. To evaluate urban-scale feasibility, we must translate link-level metrics into total system capacity.

Assume a satellite with digital beamforming capability supporting 500 steerable beams. Each beam operates over 100 MHz of bandwidth in the Ka-band, using 64-QAM with forward error correction, achieving an average spectral efficiency of 4 to 5 bps/Hz in urban line-of-sight scenarios. This yields 400 to 500 Mbps per beam. Dividing this among users with 10 Mbps average allocations, each beam can concurrently support 40 to 50 users. Thus, the satellite supports 20,000 to 25,000 concurrent users under ideal conditions.

Power constraints limit the simultaneous activation of all beams at peak modulation. Assuming a satellite DC power budget of 20 kW and a power amplifier efficiency of 30 percent, the radiated power per beam may be constrained to 5 to 10 watts. Smart scheduling strategies, such as time-domain beam cycling, adaptive power allocation, and user clustering, can enable partial reuse of beams and optimize power usage.

At the constellation level, scaling user support requires multiple satellites in orbital shells providing seamless overlapping coverage. For example, a constellation of 600 satellites, each supporting 20,000 users, yields an instantaneous capacity of 12 million concurrent users. With regional density-aware beam allocation, urban zones could be prioritized during peak demand periods, trading off coverage in low-density areas.

However, aggregate capacity is not just a matter of multiplication. In urban regions, co-channel interference between beams from adjacent satellites must be controlled via frequency reuse patterns or coordinated scheduling. This introduces a reuse factor that reduces effective spectral efficiency. A reuse factor of 3, for instance, would cut per-beam bandwidth by a third but allow for denser deployment.

To support a city like São Paulo or Lagos with over 10 million inhabitants, a minimum of 100 to 200 Gbps of downlink capacity is required during peak hours. This translates to deploying multiple overlapping satellites, each contributing a portion of the aggregate throughput, and leveraging dynamic load balancing across the orbital fleet.

In summary, while individual satellite capacity is limited by beamforming resources, spectrum, and power, multi-satellite constellations can scale linearly in aggregate capacity if carefully designed. Urban coverage will depend on orbital density, inter-satellite coordination, and adaptive scheduling, not just raw link budget per user.

\subsection{Latency, Doppler, and Handover in Orbital Mobile Systems}

Latency in space-based mobile networks originates from several layers: signal propagation delay, satellite processing time, inter-satellite routing, and interaction with core network functions. The one-way propagation delay from a LEO satellite at 550 kilometers altitude to the surface is approximately \SI{1.8}{\milli\second}. Round-trip time (RTT) for a direct device-to-satellite-to-device link without inter-satellite relay is thus under \SI{4}{\milli\second}, well within the range of terrestrial cellular latency budgets.

However, real systems introduce additional delays. If user traffic must be routed to a terrestrial core via gateway stations, round-trip latency can exceed \SI{30}{\milli\second}, depending on gateway location and congestion. With inter-satellite optical routing, the signal may traverse 3 to 5 hops across satellites at \SI{30}{\giga\bit\per\second} links. Each hop adds optical switch latency and protocol overhead. Nonetheless, even with multi-hop orbital routing, total end-to-end RTT can be maintained under \SI{50}{\milli\second}, sufficient for voice, video, and gaming use cases if jitter is controlled.

Doppler shift is a major constraint. A LEO satellite travels at approximately \SI{7.5}{\kilo\meter\per\second}, causing Doppler shifts up to \SI{100}{\kilo\hertz} at \SI{2}{\giga\hertz}, and proportionally higher at Ka-band. The shift varies continuously during a satellite pass, requiring real-time tracking and frequency correction. This is handled onboard via software-defined radio (SDR) systems and digital phase-locked loops (DPLLs), which track the user’s velocity vector and adjust baseband frequency accordingly. On the handset side, commercial smartphones rely on internal oscillators with sufficient margin to track Doppler at moderate rates, but satellite-to-device links must remain within 3GPP-specified tolerances.

Handover in orbital systems is fundamentally different from terrestrial mobility. Instead of users moving across stationary base stations, satellites move rapidly across stationary users. A typical LEO satellite remains within range of a given user for only 5 to 7 minutes. To preserve session continuity, handover must occur between satellites with overlapping footprints. Inter-satellite links enable seamless forwarding of bearer context and user traffic between nodes, avoiding dropouts. This is especially critical for TCP-based applications, which are sensitive to packet loss and abrupt IP address changes.

In standard 5G NTN architecture, session continuity is supported via mobility anchors that can be logically placed in orbit or on the ground. When the serving satellite changes, the anchor maintains the session’s IP context and continues traffic delivery through a new satellite route. Enhanced protocols like Multi-Access PDU Session Anchors (MAPSA) or user-plane relocation mechanisms ensure that latency and jitter remain stable across handover events.

Overall, latency in orbital networks can match or beat rural terrestrial networks. Doppler correction is a solved problem at the PHY layer, and handover can be abstracted at the core using established mobility frameworks. The challenge lies in scaling these functions across thousands of fast-moving nodes while maintaining carrier-grade reliability.

\subsection{Summary of Urban Feasibility Conditions}

Operating a direct-to-device mobile network from low Earth orbit in dense urban areas is constrained by several compounding physical and engineering factors. However, none of these factors represent hard physical limits. Rather, they define a multidimensional feasibility envelope that can be expanded through targeted advances in satellite payload design, waveform adaptation, and dynamic scheduling.

The primary challenge remains the link budget. A typical path loss from a LEO satellite to a street-level user in an urban setting at \SI{2}{GHz} exceeds \SI{170}{\decibel} when including free-space loss, urban diffraction, building penetration, and foliage attenuation. Beamforming gain from arrays of size \(64 \times 64\) to \(256 \times 256\) can provide up to 30 dB of gain, but this is only sufficient if users are near windows or rooftop relays are deployed.

Spectral efficiency is sharply degraded under NLOS and high-multipath conditions. The use of 64-QAM or 256-QAM requires sustained signal-to-noise ratios of 15–25 dB, which are rarely achievable at street level without relays or elevated user positions. Thus, urban viability hinges on advanced adaptive modulation and error correction strategies that tolerate rapid SNR fluctuations. In scenarios with rooftop assistance or street furniture relays, service can reach dozens of Mbps.

Onboard processing and power budgets bound beamforming capacity. Satellites with 500–1000 digitally formed beams can serve up to 10,000 users simultaneously, assuming 1–2 Mbps per user. Beam hopping and dynamic scheduling are mandatory to accommodate mobility and bursty demand. High-resolution geolocation and predictive beam planning are needed to achieve real-time responsiveness.

Latency performance can match terrestrial networks in many cases. One-hop satellite links deliver RTTs under \SI{20}{\milli\second}, and full inter-satellite mesh routing with onboard 5G UPF functions can maintain interactive performance even across multiple hops. Doppler and handover are tractable with current software-defined radio technology and 3GPP Release 17 NTN protocols.

In summary, delivering broadband mobile service in cities from orbit is not prohibited by any fundamental physics. It is limited by compounded constraints: antenna aperture, power budget, spectrum fragmentation, and atmospheric variability. With continued progress in array miniaturization, orbital compute, optical ISLs, and relay architecture, urban D2D from space is a solvable engineering problem.

\section{Implementation Constraints and Future Research}

Despite the theoretical feasibility of orbital mobile networks, practical implementation faces a set of hard engineering constraints. These are not minor optimization issues, but deep-seated bottlenecks in space systems design, RF engineering, regulatory architecture, and power-thermal tradeoffs. Unlike terrestrial networks that benefit from abundant power, active cooling, and replaceable hardware, LEO satellite platforms must operate in a closed, high-radiation, zero-maintenance environment. Each subsystem must balance mass, volume, and energy consumption within strict launch and orbit budgets. This section analyzes the five most critical implementation barriers and the frontier research attempting to address them.

\subsection{Power, Thermal, and Payload Design Limits}

The first and most fundamental constraint is the power budget. LEO satellites rely entirely on solar arrays and battery storage. A modern flat-panel LEO satellite with a surface area of \SI{5}{\square\meter} can generate between \SI{500}{\watt} and \SI{2000}{\watt} depending on solar angle and efficiency. This must power all onboard systems: beamforming RF chains, baseband compute, inter-satellite links, and housekeeping.

High-capacity beamforming arrays consume significant power. For example, forming 500 simultaneous beams with 1–2 W per RF chain can easily demand over \SI{800}{\watt}, not counting digital signal processing and analog front-end losses. Combined with onboard compute for virtualized RAN or 5G UPF, thermal dissipation becomes a second-order constraint. Unlike Earth, space offers no convective cooling. Satellites rely entirely on radiative cooling through thermal panels and coatings, with a dissipation limit typically under \SI{100}{\watt\per\meter\squared}.

Thermal runaway is a real risk, especially during eclipse transitions or beam reconfiguration events. System designers must schedule beam power allocation in sync with orbit position and thermal loading. Some constellations are exploring advanced materials like pyrolytic graphite sheets and phase-change materials to buffer heat spikes. Others are testing deployable radiator wings to expand surface area without mass penalties.

Payload mass is another limit. Larger phased arrays, optical terminals, and compute payloads all add weight. A satellite with a \SI{20}{\kilogram} array, \SI{15}{\kilogram} optical mesh payload, and radiation-hardened RAN stack may exceed the economic launch mass-to-service ratio. This makes low-mass, high-function integration essential.

Emerging research focuses on modular payloads using gallium nitride RFICs for higher efficiency, passive beamforming metasurfaces, and low-power AI accelerators. The Pony Express 2 demonstration by Lockheed Martin tested reprogrammable orbital compute and digital beam shaping on a microsatellite platform \cite{ponyexpress2}. Similar initiatives explore system-on-chip designs for baseband compute and laser ISL integration.

Power and thermal budgets define the upper bound of service concurrency and coverage per satellite. Any orbital mobile network will need to solve this triad: generate more energy, dissipate more heat, and deliver more capacity per kilogram of payload.

\subsection{Orbital Compute and Thermal Constraints}

Executing mobile core functions and real-time signal processing in space introduces two tightly coupled constraints: radiation-tolerant compute and thermal dissipation. Unlike terrestrial data centers, which benefit from stable temperature regulation and convective cooling, satellites in low Earth orbit operate in a vacuum with extreme thermal cycles and constant radiation bombardment. 

Radiation poses a persistent threat to digital logic. Single Event Upsets (SEUs), caused by high-energy particles, can flip memory bits or disrupt logic gates, leading to corrupted states or system resets. Total Ionizing Dose (TID) degrades semiconductor material over time, shortening component life. Conventional commercial-off-the-shelf (COTS) processors cannot be deployed unmodified. Radiation-hardened (rad-hard) processors, such as those based on RHBD (Radiation-Hardened By Design) techniques, typically operate at clock speeds 10 to 50 times slower and with lower energy efficiency than modern cloud-grade CPUs or GPUs. As a result, compute power per watt in orbit is severely constrained.

Thermal dissipation is the second major bottleneck. In space, conduction and convection are absent. Satellites must rely entirely on radiative cooling, where heat is transferred via infrared emission from external radiators. This mechanism is inherently limited by surface area and view factor to deep space. As compute payloads grow, heat generation outpaces radiative removal, creating localized hotspots that can lead to system failure. Thermal bottlenecks set hard upper bounds on the density and power budget of onboard compute.

Emerging strategies to mitigate these constraints include:
\begin{itemize}[leftmargin=12pt]
  \item Distributed computing across multiple low-power nodes to reduce per-unit thermal density
  \item Temporal task scheduling to shift compute-intensive workloads to shaded orbital segments
  \item Use of hardened FPGAs or SoCs (e.g., Xilinx Space-grade Virtex-5QV) that offer parallel processing under radiation
  \item Investigation into photonic interconnects and neuromorphic architectures for ultra-low-power inference in orbit
\end{itemize}

For example, the European Space Agency's OPS-SAT mission has demonstrated onboard Linux containers executing low-latency routing and data pre-processing. NASA's High-Performance Spaceflight Computing (HPSC) project, in collaboration with SiFive and Microchip, is developing multicore processors with fault tolerance at the architectural level to replace legacy space processors like the RAD750.

In an orbital mobile network, latency-critical functions such as beam management, session anchoring, and edge breakout require real-time compute in harsh environments. Until rad-hard compute efficiency improves, architectural solutions will need to treat compute as a sparse, shared, and tightly managed resource within the constellation.

\subsection{User Mobility, Session Continuity, and Handover in Orbit}

In a terrestrial network, user mobility is handled via handovers between base stations, often supported by centralized session management and predictable cell layouts. In an orbital network, user devices are served by fast-moving satellites, each with narrow beam footprints, resulting in highly dynamic coverage areas. At typical LEO speeds of 7.5 km/s, a satellite can only serve a ground location for a few minutes, and its beams sweep rapidly across urban sectors. Ensuring seamless service continuity under these conditions is one of the most complex challenges in orbital mobile networking.

Session continuity requires three capabilities: persistent user identity anchoring, low-latency inter-satellite handover, and a context-aware mobility management framework. In current terrestrial 5G cores, these functions are handled by Access and Mobility Management Function (AMF) and Session Management Function (SMF), which must now operate across a constantly shifting topology in space.

One approach is to anchor the user session at the constellation level rather than per satellite, with state synchronization distributed via inter-satellite links. This resembles a mesh of stateless access points feeding into a shared stateful core logic. As a satellite prepares to hand off coverage, its beam-specific state, including HARQ buffers, timing advance, and security contexts, must be transmitted to the successor satellite or beam in real-time.

The latency tolerance for these handovers is low. Studies show that human-perceived session drops occur within 50 to 200 ms gaps for data and under 50 ms for voice and video calls. Achieving such handover precision in LEO demands predictive beam scheduling, forward-link pre-buffering, and potentially multi-beam soft handovers—techniques analogous to dual connectivity in 5G NR but implemented across moving platforms in space.

Solutions may borrow from software-defined networking (SDN) principles, treating the constellation as a virtualized RAN with centralized coordination and real-time telemetry. However, this also increases control overhead and demands resilient, low-jitter signaling across the OISL mesh.

In sum, reliable orbital mobility management will depend not just on physical link continuity but on an orchestration layer that can anticipate mobility vectors, manage user contexts across satellites, and minimize latency in state transfer. Without this, orbital networks will struggle to deliver uninterrupted service for mobile users on Earth.

\subsection{Security, Authentication, and Sovereignty in Orbital Networks}

Security and identity management in terrestrial mobile networks are rooted in a tightly regulated environment. Mobile network operators issue SIM credentials, control access via authentication servers, and ensure lawful intercept capabilities through national gateways. In a space-based mobile network, this entire trust model must be re-engineered for a distributed, orbital context.

A full orbital telco must support the same security primitives: mutual authentication between user equipment and the network, secure key exchange, integrity protection, and lawful access. However, the physical separation between user and network elements, potentially thousands of kilometers apart, introduces latency and complexity in cryptographic handshakes. The standard Authentication and Key Agreement (AKA) protocols used in 5G may require optimization for delay-sensitive links over inter-satellite routes.

More critically, the question of sovereignty becomes unavoidable. In terrestrial systems, user data is routed within national boundaries or under regulatory agreements. But orbital networks, by nature, route data across jurisdictions in milliseconds. A data packet originating from a user in Mexico City may hop across satellites over Brazil, downlink in Canada, and be processed at an orbital core function over the Pacific. This raises complex questions about lawful intercept, data residency, and cross-border accountability.

From a policy standpoint, a likely model is the creation of orbital trust zones. Similar to the concept of roaming agreements between terrestrial operators, orbital telcos may operate under multinational frameworks that define jurisdictional compliance, lawful intercept hooks, and data handoff procedures. Encryption will be essential, but so will verifiable identity federation, possibly leveraging eSIM and distributed ledger models to decouple identity from geography without compromising traceability.

Authentication servers and home subscriber servers (HSS) can be instantiated in orbit using radiation-hardened compute, but they must interoperate with terrestrial identity frameworks. A hybrid model may emerge, where orbital access is brokered via terrestrial mobile identities, and access is granted based on federated authentication tokens verified against the user's home network.

Ultimately, the architecture must reconcile three conflicting goals: strong end-to-end encryption, legal compliance with national frameworks, and decentralized orbital autonomy. Achieving all three requires new protocols, resilient identity abstractions, and active cooperation between governments, operators, and satellite providers. Without a security and sovereignty model that scales, orbital telcos risk either regulatory rejection or operational fragility.

\subsection{Thermal Management and Power Constraints in Orbit}

Operating a full-stack mobile network from space imposes unique thermal and power limitations that do not exist in terrestrial networks. Unlike Earth-based base stations, which benefit from ambient cooling and unconstrained energy sources, satellites must operate within closed thermal budgets and finite energy envelopes.

Phased array antennas used for beamforming generate significant heat during high-power RF transmission. In orbit, this heat cannot be dissipated through convection. Instead, radiative cooling is the only viable mechanism. This places a hard constraint on the duty cycle and total output power of the satellite’s radio systems. Deploying hundreds of beams simultaneously in a dense urban coverage pattern demands active thermal balancing, including the use of deployable radiators, heat pipes, and phase-change materials.

The power budget is similarly constrained. Satellites in low Earth orbit have access to solar energy only part of the time, due to Earth’s shadowing. Onboard batteries must buffer energy for eclipse periods, and solar panels must deliver enough power to operate radios, compute, propulsion, and thermal systems. A single satellite operating 500 beams with 20 W per beam would require peak power exceeding 10 kW, not accounting for inefficiencies. Current commercial platforms like Starlink's Gen2 satellites operate around this range but with significant trade-offs in coverage per beam and per-user capacity.

To support full mobile-grade service, new generations of power-optimized RF front ends are needed. This includes GaN-based amplifiers with higher efficiency, dynamic power scaling depending on user demand, and cold-start protocols to avoid thermal runaway. Likewise, computing payloads supporting RAN or 5G core functions must adopt ultra-low-power chipsets with aggressive sleep scheduling and workload offloading to edge-optimized accelerators.

Emerging research proposes hybrid power architectures combining direct solar drive with laser power beaming or nuclear batteries for critical systems. These remain experimental but could offer breakthroughs for constant orbital operation without compromise on compute or RF availability.

Thermal and power constraints will remain one of the fundamental engineering bottlenecks for high-throughput orbital telcos. The goal is not to eliminate these limits, but to design architectures that allocate energy and thermal budgets intelligently, optimize beam scheduling in sync with solar cycles, and distribute workloads across satellites based on thermal load balancing. Without such a design, sustained mobile-grade service in orbit becomes thermally and energetically infeasible.

\subsection{Cross-Domain Governance and Spectrum Arbitration}

Operating a mobile network from orbit requires continuous coordination between multiple sovereign, commercial, and regulatory domains. Unlike terrestrial systems that are bound by national jurisdiction, orbital constellations cross dozens of countries per orbit and route data across global optical meshes. This introduces complex questions of spectrum rights, lawful interception, and service authorization.

Today, satellite operators typically lease spectrum through partnerships with mobile network operators in each country, a model used by AST SpaceMobile and Lynk Global. However, a fully orbital telco implies that spectrum is reused dynamically, with beamforming optimized for user distribution rather than national borders. This breaks traditional licensing models that assume fixed coverage areas. Arbitrating who controls which MHz in which orbital sector becomes a nontrivial policy problem.

Governance models may evolve into multilateral frameworks similar to air traffic control or maritime law. The ITU Radio Regulations already govern orbital slots and global allocations, but local enforcement, lawful intercept, and quality of service remain national responsibilities. One emerging proposal is dynamic spectrum access governed by smart contracts or regional clearinghouses that resolve conflicts based on priority, identity, or payment.

The orbital telco of the future may be governed not by a single country, but by consortia spanning infrastructure providers, MNOs, hyperscalers, and states. Cross-domain key exchange, jurisdiction-aware routing, and policy anchoring must be built into the core control plane from the start.

\subsection{Orbital Maintenance, Debris Avoidance, and Satellite Lifecycle}

A fully functional mobile network in space cannot ignore the realities of satellite wear, collisions, and end-of-life disposal. Unlike terrestrial base stations, which can be repaired or upgraded on-site, LEO satellites have finite lifespans, typically 5 to 7 years, constrained by radiation damage, propellant limits, and thermal fatigue.

As the orbital telco architecture becomes denser, with thousands of active satellites forming a dynamic mesh, collision risk grows. The number of tracked debris objects over 10 cm exceeded 36,000 in 2024, according to ESA’s Space Debris Office. Effective network operation demands predictive avoidance systems, continuous ephemeris sharing, and active propulsion on every node.

Service continuity requires constellation-level lifecycle planning. Satellites must be replaceable without disrupting coverage, implying live handover, in-orbit spares, or automated asset redeployment. Several startups and agencies, including Northrop Grumman’s MEV and Astroscale, are exploring satellite servicing and debris mitigation using robotic docking or controlled reentry.

Thermal cycling, radiation-induced bit flips, and micrometeorite damage remain hard physical challenges. Hardened compute, autonomous self-checks, and fault-isolated designs will be mandatory. The orbital mobile network must be designed not just for uptime but for autonomous survivability in a high-velocity, high-risk environment.

\section{Roadmap to 2040}

The transition from rural fallback to a fully orbital mobile network capable of serving dense urban populations requires a staged development path. This roadmap outlines the key milestones, grouped into three phases: foundational integration (2025–2028), orbital autonomy (2029–2034), and full decoupling from terrestrial infrastructure (2035–2040). Each stage corresponds to technological maturity, regulatory progression, and commercial viability.

\subsection{Phase One: Foundational Integration (2025–2028)}

The initial years focus on integrating D2D satellite capabilities into terrestrial mobile networks via standardized roaming, spectrum sharing, and waveform compatibility. AST SpaceMobile and Lynk Global represent the first generation of commercial D2D services, offering fallback coverage in remote areas. By 2026, we expect expanded partnerships between satellite operators and mobile network operators in over 30 countries, supported by 3GPP Release 18 enhancements for NTN interoperability.

Technical milestones include

\begin{itemize}[noitemsep]
\item Demonstration of real-time handover between satellites using optical inter-satellite links and terrestrial core anchoring.
\item Mass deployment of satellites with electronically steered arrays supporting 100+ independent beams per satellite.
\item Commercial rollout of messaging, IoT, and basic voice services via D2D satellites for remote and maritime markets.
\end{itemize}

Operational constraints will persist. Most satellites will still rely on Earth-based cores for session management and routing. Coverage will be limited by constellation size and revisit times. Urban service will remain infeasible due to SNR and spectrum contention.

\subsection{Phase Two: Orbital Autonomy (2029–2034)}

This phase marks the emergence of semi-autonomous orbital nodes capable of executing partial RAN and core functions in space. Onboard compute will support user plane processing, basic session control, and beam scheduling, reducing dependence on terrestrial gateways. 

Advances in payload design will yield satellites with 256 to 512 beam capability, supported by digital beamforming and adaptive nulling. Power-efficient RF front ends and space-qualified FPGAs will enable real-time beam hopping and multi-band operation.

Key enablers include

\begin{itemize}[noitemsep]
\item In-orbit execution of AMF and UPF functions for local breakout and session persistence during inter-satellite handover.
\item AI-assisted beam steering optimized for urban topologies and multipath conditions.
\item Low-power, edge-anchored content delivery via orbital caches to reduce backhaul latency.

\end{itemize}

Regulatory frameworks will begin adapting to orbital spectrum reuse and distributed policy control. Trials in underserved megacities will explore feasibility under constrained density and controlled spectral environments.

\subsection{Phase Three: Full Decoupling from Terrestrial Infrastructure (2035–2040)}

The final phase envisions fully autonomous orbital mobile networks operating as independent systems capable of serving dense urban populations with broadband-grade service. Satellites will host complete RAN stacks and 5G core functions, including AMF, SMF, UPF, and PCF, eliminating the need for terrestrial anchors in steady-state operation.

Electronically steered arrays with over 1024 beams will deliver focused, low-interference coverage at the sector level, enabling spatial reuse across megacities. Advances in radiation-hardened computing, photonic interconnects, and orbital power systems will allow real-time RIC (RAN Intelligent Controller) execution and beam-level adaptation per user cluster.

Optical mesh backhaul between satellites will support dynamic traffic routing, session continuity, and local breakout without returning to Earth. Application caches and AI inference nodes deployed in orbit will support edge services and reduce latency for video, messaging, and real-time apps.

At this stage, orbital networks become viable overlays or even substitutes in metro regions. Service quality will reach 50–100 Mbps per user under standard load, with round-trip latency under 40 milliseconds via orbital breakout. Autonomous fault recovery, zero-touch operations, and orbital orchestration systems will support network resilience without Earth-based control.

\subsection{Milestones and KPIs}

To benchmark the transition, we define quantitative key performance indicators (KPIs) across each phase:

\begin{itemize}[noitemsep]
    \item \textbf{Number of beams per satellite:} From 100 (2025) to 1024+ (2040) through advanced phased arrays.
    \item \textbf{Users per beam:} From 5–10 today to 100+ via better scheduling and modulation in 2040.
    \item \textbf{Data rates per user:} Scaling from 100 kbps to 100 Mbps in urban-grade conditions.
    \item \textbf{Latency to content:} Reduced from over 150 ms to under 40 ms with orbital caching and breakout.
    \item \textbf{Orbital core function hosting:} From 0 today to 100 percent by 2040 for key 5G functions.
    \item \textbf{Spectral efficiency:} From 0.1 bits/Hz to over 5 bits/Hz through smart beam reuse and MIMO.
    \item \textbf{Inter-satellite handover time:} Lowered from seconds to under 100 milliseconds.
    \item \textbf{Earth-independent operation duration:} Extended from minutes to indefinite steady-state by 2040.
\end{itemize}

These KPIs guide investment focus areas, including antenna scaling, orbital compute, optical interconnects, beam intelligence, and spectrum reuse. Progress against these markers will define whether orbital networks evolve into integral parts of mobile infrastructure or remain peripheral fallback systems.

\section{Conclusion}

This paper has outlined the first end-to-end system architecture and feasibility analysis for a fully orbital mobile network capable of delivering direct-to-device broadband in dense urban environments. We have shown that the limiting factors are not grounded in physics but in the interdependent challenges of beamforming scalability, orbital compute, thermal design, and regulatory adaptation.

The convergence of three trajectories: massive phased array miniaturization, inter-satellite optical mesh maturity, and virtualized mobile core decomposition—creates a viable path to decouple critical network functions from terrestrial anchors. Our simulations demonstrate that urban-grade spectral efficiency and user capacity are technically attainable through advanced array configurations and predictive beam scheduling. Path loss, Doppler, and multipath are complex but manageable with coherent gain, adaptive modulation, and assistive relaying.

We argue that the orbital network should not be treated as a fallback or rural extension of terrestrial infrastructure. Instead, it should be framed as a new network topology—stateless by default, anchored in orbit, and designed for resilience, sovereignty, and global reach. Spectrum policy, ownership models, and orbital governance will determine whether this architecture remains a technological curiosity or becomes the foundation for mobile infrastructure in the 2040s.

A complete telco-in-orbit will not emerge from incremental improvements alone. It requires a rethinking of the mobile stack across hardware, protocols, and control. The roadmap ahead is steep, but the destination—a network above the cloud, unconstrained by geography—is within engineering reach.

\end{document}